# Comparison of the Scaling Properties of EUV Intensity Fluctuations in Coronal Hole and Quiet-Sun Regions

Ana Cristina Cadavid,[1] Mari Paz Miralles,[2] and Kristine Romich[1]

[1]*Department of Physics and Astronomy*
*California State University Northridge*
*18111 Nordhoff Street*
*Northridge, CA 91330, USA*
[2]*Center for Astrophysics | Harvard & Smithsonian*
*60 Garden Street*
*Cambridge, MA 02138*

## ABSTRACT

Using detrended fluctuation analysis (DFA) and rescaled range (R/S) analysis, we investigate the scaling properties of EUV intensity fluctuations of low-latitude coronal holes (CHs) and neighboring quiet-Sun (QS) regions in signals obtained with the Solar Dynamics Observatory/Atmospheric Imaging Assembly (SDO/AIA) instrument. Contemporaneous line-of-sight SDO/Helioseismic and Magnetic Imager (HMI) magnetic fields provide a context for the physical environment. We find that the intensity fluctuations in the time series of EUV images present at each spatial point a scaling symmetry over the range $\sim$20 min to $\sim$1 hour. Thus we are able to calculate a generalized Hurst exponent and produce image maps, not of physical quantities like intensity or temperature, but of a single dynamical parameter that sums up the statistical nature of the intensity fluctuations at each pixel. In quiet-Sun (QS) regions and in coronal holes (CHs) with magnetic bipoles, the scaling exponent ($1.0 < \alpha \leq 1.5$) corresponds to anti-correlated turbulent-like processes. In coronal holes, and in quiet-Sun regions primarily associated with (open) magnetic field of dominant polarity, the generalized exponent ($0.5 < \alpha < 1$) corresponds to positively-correlated (persistent) processes. We identify a tendency for $\alpha \sim 1$ near coronal hole boundaries and in other regions in which open and closed magnetic fields are in proximity. This is a signature of an underlying $1/f$ type process that is characteristic for self-organized criticality and shot-noise models.

## 1. INTRODUCTION

Corresponding author: Ana Cristina Cadavid

Coronal holes (CHs) are characterized by having low density, a greatly reduced extreme ultraviolet emission at coronal temperatures, and a predominance of open magnetic flux as compared to their neighboring regions (e.g., Cranmer 2009, and references therein). While there is general acceptance that the fast solar wind primarily originates in polar coronal holes (Wilcox 1968; Krieger et al. 1973; Cranmer et al. 2017), in the case of the slow solar wind, the situation is more complex, with different magnetic structures and physical processes involved (Abbo et al. 2016, and references therein). In this context, an area of research has centered on understanding the properties of low-latitude CHs (Miralles et al. 2004, 2006) and neighboring quiet-Sun (QS) regions, along with the role they play in coronal heating and solar wind acceleration. The difference in coronal emission between CHs and the adjacent QS is not apparent in the chromosphere and transition region (Wiegelman & Solanki 2004, and references therein). After performing a statistical analysis of the magnetic properties, these authors concluded that low-latitude CHs and QS regions have comparable numbers of short closed loops, leading to similar radiation at chromospheric temperatures. In contrast, at coronal heights the longer loops common in QS are rare in CHs. Instead, these regions are characterized by open magnetic field lines which allow outward flow of the plasma. Further studies have considered the interactions between the open and closed fields and a general picture has emerged in which bipole structures are advected to the boundary of supergranular cells where they reconnect with the open fields. In CH open flux funnels, the energy is released to accelerate outflows, while in regions with overlying closed loops (as in the QS), the energy is released into the loop structures (Tu et al. 2005; McIntosh et al. 2007; He et al. 2010). It has also been shown that the interchange reconnection between open and closed fields at the CH boundaries can lead to plasma emission in jet-like events (Subramanian et al. 2010), and allow the boundaries to evolve so that the CHs can maintain their effective rigid rotation (Sheeley et al. 1987; Madjarska & Wiegelmann 2009; Yang et al. 2011; Tian et al. 2011).

The solar atmosphere presents time series which combine periodic and quasiperiodic signals, trends, and stochastic signals (e.g., De Moortel & Nakariakov 2012; Uritsky et al. 2013; Arregui 2015). Using data from the Atmospheric Imaging Assembly (AIA) and the Helioseismic and Magnetic Imager (HMI) instruments on board the Solar Dynamics Observatory (SDO), we have investigated the scaling properties of the stochastic component in time series of EUV emission in low-latitude CHs and neighboring QS regions in the context of the magnetic field topology. Previously, Cadavid et al. (2014) and Ireland et al. (2015) found that AIA EUV emission in different coronal regions presents Fourier power spectra with spectral scaling exponents $\beta > 1$, rather than white noise with $\beta = 0$, indicating that the time series are statistically nonstationary (Mandelbrot & Van Ness 1968; Davis et al. 1994). In a standard working definition, a stationary stochastic process is characterized by an autocorrelation which depends only on the differences of time intervals, while in the nonstationary case some of the statistical properties are deterministic functions of time (eg. Koutsoyiannis 2011). Nonstationarities may arise from external effects, such as long-term trends or externally driven periodic signals, in which case it is important to remove them.

Alternatively the nonstationarities can be an intrinsic component of the process (such as long-term correlations), in which case they must be examined in detail. For this purpose we turn to the method of detrended fluctuation analysis (DFA), which was designed precisely to determine the true scaling properties of a signal by identifying long-term correlations in noisy and nonstationary time series after accounting for external trends (i.e., Peng et al. 1994; Chen et al. 2002; Hu et al. 2001; Kantelhardt et al. 2002). It has subsequently been shown that DFA can also be used in the analysis of stationary time series, and can therefore be applied without previous knowledge of the time series' statistical properties.

Since our goal is to investigate the scaling symmetries for individual pixels, DFA proves to be a useful tool and leads to a more accurate identification of scaling exponents as compared to traditional power spectra analysis. Using DFA, we are able to calculate a generalized Hurst exponent $\alpha$ and produce image maps, not of physical parameters like intensity or temperature, but of this single dynamical parameter that sums up the statistical nature of the intensity fluctuations at each pixel. The values of the generalized Hurst exponent permit us to quantify the differences between CHs, neighboring QS regions, and the boundaries between them, and identify the characteristic properties of the long-term variability of the underlying physical processes. In turn, these are related to the magnetic structures in contemporaneous SDO/HMI magnetic images and corresponding potential field extrapolations of the photospheric magnetic fields.

The original work on the long-term variability of stationary time series was presented by Hurst (1951, 1957) and the actual statistical description was developed later by Mandelbrot and Van Ness (1968). By using what is now known as rescaled range (R/S) analysis, Hurst identified the scaling properties of different geophysical time series and quantified their long-term variability via the "Hurst coefficient" (or exponent) $H$. It has been shown that DFA can be used to estimate the Hurst coefficient through a relation established with the generalized Hurst exponent $\alpha$ (i.e. Heneghan & McDarby 2000). In this work, we also apply the R/S analysis technique as a complement to DFA, since in particular it tends to give a more accurate estimate of the scaling properties of the time signals in CH boundaries. The Hurst exponent can be directly related to the more widely-used power spectra scaling exponent, so we will further interpret the results obtained for the CH and QS regions in terms of processes with power spectra of the form $S(f) \sim 1/f^{\beta}$. This information will aid in identifying the possible origins of the local scaling properties. Of particular interest is the $1/f$ noise ($\beta \sim 1$) process, which has been identified in a wide variety of systems, including the magnetic field fluctuations in the solar wind (Bruno & Carbone 2013, and references therein).

The paper is organized as follows: In Section 2, we describe the data and the techniques used to identify CH and adjacent QS regions. In Section 3, we introduce the mathematical details of DFA and R/S analysis and the practical considerations that arise when these methods are applied to observational data. In Section 4, we report the results for three different coronal holes and surrounding QS regions, analyzing the $\alpha$ and $H$ maps in the context of the line-of-sight magnetic field images and their potential field extrapolations.

Finally, in Section 5, we summarize our findings and further discuss the implication of the results.

## 2. DATA AND DATA PREPARATION

In this study, we use data from the AIA instrument (Lemen et al. 2012) on board SDO (Pesnell et al. 2012), which is in virtually continuous operation and covers the full solar disk with a $0.6''$ ($\sim$0.44 Mm) pixel scale (spatial resolution of $1.2''$ or $\sim$0.9 Mm). We focus on two extreme ultraviolet (EUV) channels (171 Å and 193 Å), which cover the upper transition region and lower corona, respectively. The time series have a duration of 5 hours, which at AIA's 12-second cadence consists of 1500 temporal pixels; this is a sufficient number of data points to implement the scaling analyses. The data cubes were obtained using the Joint Science Operation Center (JSOC) Cutout Service selecting the "tracking" option which accounts for the solar rotation by tracking the image patch at the Carrington rate. In addition the images were coaligned by using the central image in the time series as a reference. Contemporaneous line-of-sight (LOS) images from HMI (Schou et al. 2012), together with potential field extrapolations, were used to establish the magnetic context. We present results for three coronal holes and adjacent QS inserted within $400'' \times 400''$ regions. The data corresponding to the first coronal hole (CH1) was taken on 2017 April 16 starting at 12:00:05 UT with the region centered at (0, -50''). The corresponding dates, starting times, and region centers for the other two coronal holes are: CH2 – 2017 April 21 12:00:00 UT (0, 50''); CH3 – 2017 February 27 00:00:05 UT (160'', -350'').

To define a general boundary for the CHs throughout the time series, we modify the method of Krista & Gallagher (2009) as follows. We consider the histogram of all 193 Å intensity values in the data block of time series and identify peak values for the CH and QS intensities with a minimum in between. This value is chosen as the threshold to define the coronal hole boundary. Qualitative comparison is made with the contour presented for the corresponding image in Helioviewer (https://www.helioviewer.org) based on the Spatial Possibilistic Clustering Algorithm (SPOCA; Verbeeck et al. 2014). Figure 1 displays the logarithm of the average intensities for the 193 Å (top row) and the 171 Å (bottom row) time series for the three low-latitude coronal holes. The superimposed boundaries are obtained from the thresholds in the 193 Å data. Coronal hole CH2 can also be distinguished in the 171 Å intensity average, but this is not the case for CH1 and CH3.

## 3. ANALYSIS METHODS

### 3.1. *The Hurst and Spectral Exponents*

Spectra obeying a power law are encountered in a variety of astrophysical systems (Aschwanden 2011, and references therein). Recent examples for observations in the solar atmosphere include results for coronal emission (e.g. Auchère et al. 2014; Gupta et al. 2014; Cadavid et al. 2014; Ireland et al. 2015) and chromospheric signals (e.g. Reardon et al. 2008; Lawrence et al. 2011; Krishna Prasad et al. 2017). The motivation to use DFA in this study is based on two points: first, the fact that the time series of EUV coronal emission are nonstationary, and second, to have a technique which can quantify more precisely the scaling properties of individual pixel time series. Furthermore it is possible to relate the generalized Hurst exponent $\alpha$ obtained with DFA, the original Hurst exponent $H$, and the spectral power exponent $\beta$, allowing us to make contact with standard results.

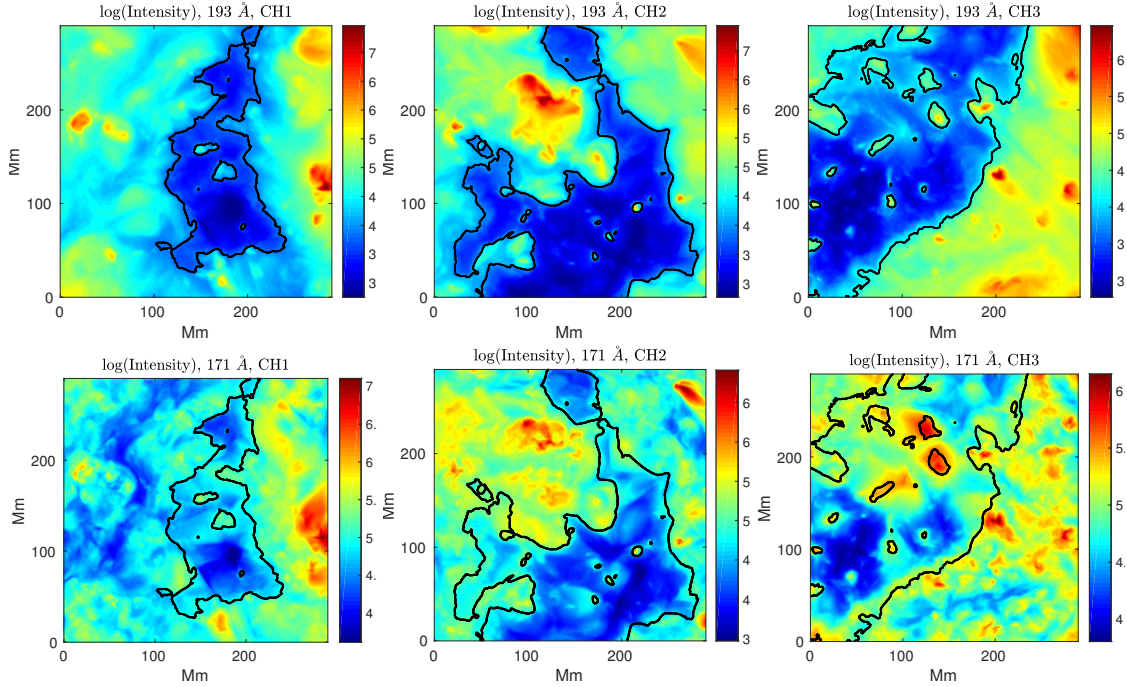

**Figure 1.** Logarithm of the intensity (in DN) for three low-latitude coronal holes and adjacent QS regions in the 193 Å (top row) and 171 Å (bottom row) emission: CH1 (2017 April 16; left), CH2 (2017 April 21; middle), CH3 (2017 February 21; right). The boundaries denoted by the black line are specified at 45 DN, 55 DN and 65 DN of the 193 Å intensity, respectively.

As it is well known, the presence of a power law in the spectra is representative of an underlying scaling symmetry. In particular, self-similarity describes the property in which a part of a physical structure replicates the whole structure upon rescaling (e.g. Feder 1988, and references therein). In the case of a time series in which the rescaling symmetry only applies to the time dimension and not to the amplitude of the signal, the self-similarity is instead a self-affinity (Hardstone 2012), which can be described in terms of the statistical properties of the time series. A stochastic nonstationary process $X(t)$ is self-affine if it satisfies

$$X(\lambda t) \equiv \lambda^H X(t),$$                                          (1)

where $\lambda$ is a scaling factor, $0 < H < 1$ is the Hurst exponent, and $\equiv$ means that the variance is the same for both sides of the equation (Feder 1988; Beran 1994; Gilmore et al. 2002; Hardstone et al. 2012). The Hurst exponent provides a measure of the long-term memory of a time series: $H = 0.5$ indicates an uncorrelated series; $H > 0.5$ corresponds to a positively-correlated ("persistent") series in which a positive (negative) fluctuation tends to be followed by a positive (negative) fluctuation; $H < 0.5$ corresponds to an anti-correlated ("antipersistent") series in which a positive fluctuation tends to be followed by a negative fluctuation and vice versa. The traditional model for a stochastic nonstationary self-affine process is fractional Brownian motion (fBm) (Mandelbrot & Van Ness 1968). It is a generalization of the well-known Brownian motion, which is a Gaussian process ($H = 0.5$) with increments that are not independent. For these nonstationary processes the power spectra obtained from the Fourier transform of the autocovariance is time-dependent and therefore not well-defined. A solution was obtained in the form of the Wigner-Ville spectrum but relaxing the nonnegativity property of the classical spectrum (Martin & Flandrin 1985). It has been shown that the time average of the Wigner-Ville spectrum leads to a power spectra of the form $S(f) \propto f^{-\beta}$ with $1 < (\beta = 2H + 1) < 3$ (Flandrin 1989; Heneghan & McDarby 2000).

The defining property of stationary processes $X(t)$ is that their statistical properties do not depend on time. A weaker form of stationarity just requires that the mean and the variance are independent of $t$ (independent of the time period over which these quantities are calculated). It follows that the self-affinity described by equation (1) is not relevant in this case, since by definition the variance does not change in time. Instead stationary processes can exhibit a type of scale invariance in which the autocorrelation function depends only on the time difference $\tau$ and satisfies $r(\tau) \propto \tau^{(2H-2)}$. In this case the power spectra is well-defined and satisfies $S(f) \propto f^{-\beta}$ with $-1 < (\beta = 2H - 1) < 1$. The traditional example of these scale-invariant processes is fractional Gaussian noise (fGn), which is a generalization of Gaussian noise ($H = 0.5$). Mandelbrot & Van Ness (1968) showed that the increments of a fBm signal are described by a fGn process. Similarly the cumulative sum of a fGn signal leads to a fBm self-affine process.

For the wide class of $1/f^{\beta}$ noises (Aschwanden 2011) the scaling symmetry is actually satisfied by the power spectra itself (Halley & Inchausti 2004):

$$S(\lambda f) = |\lambda|^{-\beta} S(f) \qquad (2)$$

It has been shown that these $1/f$ noises can be approximated by a fBm process when $\beta > 1$ and by a fGn process when $\beta < 1$ in the manner described above. However the $1/f^{\beta}$ processes are more general in that they do not require Gaussian increments and the values of the spectral exponents are not as restricted (Wornell 1993; Halley & Inchausti 2004).

In the next section we specify the relation between the Hurst and generalized Hurst exponents. We note that DFA will be applied to the time series without assuming any underlying model. The possible origin of the scaling properties will be considered in the discussion section based on the physical processes in the solar atmosphere.

### 3.2. *Detrended Fluctuation Analysis*

Detrended fluctuation analysis (DFA) was originally introduced by Peng et al. (1994) and has been used extensively to study the scaling and correlation properties of systems in different fields. In order to treat all processes in the same manner and exploit the self-affine properties, it is convenient to define the "profile" of a time series $X(k)$ of length $N$ as

$$Y(i) \equiv \sum_{k=1}^{i} (X(k) - \langle X \rangle), \text{ for } i = 1, ..., N, \tag{3}$$

where $\langle X \rangle$ is the mean of the time series. By calculating the cumulative sum there is no need to consider whether the time series is stationary or nonstationary before other analysis is performed (Hardstone et al. 2012).

After calculating the profile $Y$ for a time series of length $N$, the method implements the following steps as described by Kantelhardt et al. (2002):

The profile is divided into $N_s = \text{int}(N/s)$ non-overlapping segments of temporal length (scale) $s$, and a least-squares method is used to fit a polynomial $y_v$ of degree $m$ to each segment $v$. The fitted polynomial is then subtracted from each profile segment to obtain the variance

$$F^2(v, s) \equiv \frac{1}{s} \sum_{i=1}^{s} (Y[(v-1)s + i] - y_v(i))^2. \tag{4}$$

The fluctuation function is then given by

$$F(s) = \left(\frac{1}{N_s}\right) \left(\sum_{v=1}^{N_s} F^2(v, s)\right)^{1/2}. \tag{5}$$

If the time series has a scaling symmetry, the fluctuation function satisfies the scaling law $F(s) \sim s^{\alpha}$. The "generalized Hurst exponent" $\alpha$ is then obtained from log-log plots of $F(s)$ vs. $s$ for the appropriate scaling range. The exponent $\alpha$ can then be related to the Hurst exponent and applied to a range of time series with different statistical properties (Mandelbrot & Van Ness 1968; Davis et al. 1994). When applying the method we calculate the fluctuation function at equally spaced log(time) intervals up to temporal scale $s \sim N/3$ before the results become statistically unreliable. For stationary time series $0 < \alpha < 1$ and $H = \alpha$, while for nonstationary time series $1 < \alpha < 2$ and $H = \alpha - 1$. Well-known examples of time series are white noise ($\alpha = H = 0.5$), Brownian motion ($\alpha = 1.5, H = 0.5$), and red noise ($\alpha = H = 1$). If the spectral exponent $\beta < 3$ the time series have stationary increments and $\beta = 2\alpha - 1$ (Heneghan & McDarby 2000). Other reference values of interest to solar physics are $\beta = 5/3 \sim (\alpha = 1.33)$ for Kolmogorov-type turbulence, and $\beta = 3/2 \sim (\alpha = 1.25)$ for Kraichnan's model of Alfvén-wave turbulence (Marsch & Tu 1997, and references therein).

### 3.3. *Rescaled Range (R/S) Analysis*

We have followed Feder (1988) and Gilmore et al. (2002) in the implementation of the R/S analysis. For a time series $X(k)$ of length $N$, we define the temporal scales as: $s = N/2, N/2^2, ..., N/2^7$. The average over a temporal scale of length $s$ is given by $\langle X \rangle_s = \frac{1}{s} \sum_{j=1}^{s} X(j)$, where $j = 1$ corresponds to the start of a particular segment. In turn the cumulative time series for non-overlapping segments of duration $s$ is

$$W(i)_s = \sum_{j=1}^{i}(X(j) - \langle X \rangle_s), \text{for } i = 1, ..., s. \tag{6}$$

The aim of the R/S analysis introduced by Hurst is to investigate the scaling relation

$$\lim_{s \to \infty} \frac{R(s)}{S(s)} \propto s^H, \tag{7}$$

where $R(s)$ is the rescaled range and $S(s)$ the standard deviation for a process $X(k)$ which satisfies fGn statistics. The expressions for $R$ and $S$ as functions of the scale (lag) $s$ are:

$$R(s) = \max(0, W(1)_s, W(2)_s \ldots, W(s)_s) - \min(0, W(1)_s, W(2)_s \ldots, W(s)_s) \tag{8}$$

$$S(s) = \left( \frac{1}{s} \sum_{j=1}^{s}(X(j) - \langle X \rangle_s)^2 \right)^{1/2} \tag{9}$$

To extend the R/S analysis to time series with fBm statistics the first step consists of calculating the series of increments $X'(k) = X(k+1) - X(k)$, which satisfies fGn statistics.

The data blocks for a given CH region include time series with different statistical properties. DFA can be applied without *a priori* knowledge of these properties; the scaling parameter $\alpha$ can then be used to identify the class to which each pixel belongs. One can proceed to apply the R/S analysis with the understanding that it will lead to a correct estimate of the Hurst parameter $H$ only for pixels that correspond to a fGn-type process. Here $0 < H < 1$, while for other pixels $H$ will saturate at 1 or larger than 1, which is not defined. We will actually exploit this limitation to identify pixels with fluctuations obeying a $1/f$ scaling law.

## 4. RESULTS

### 4.1. *Time Series*

Before presenting the results of the scaling analyses it is useful to look at some characteristic time series to gain intuition on how they fit in the mathematical descriptions given in the last section. Figure 2 displays two examples of 193 Å intensity time series in the data for CH1 and QS1, together with the corresponding PDFs of intensity values. (Comparable time series are found in the other two coronal hole regions under study.) The first column (CH pixel) presents a time series with an approximately Gaussian PDF and $0.5 < \alpha = H < 1$, compatible with a stationary process. The second column (QS pixel) displays a time series with a non-Gaussian PDF and $1.0 < \alpha < 1.5$, compatible with a nonstationary process. (We note that these are just examples and pixels with $\alpha < 1$ and $\alpha > 1$ are found throughout the coronal hole and quiet-Sun regions.)

In order to select a common scaling range for all pixels in an image block we tested the scaling properties of the fluctuation function obtained from DFA for a collection of pixels in areas with different magnetic field properties. Figure 3 displays examples of log-log plots of the fluctuation function $F$ as a function of time scale, for a typical single pixel in CH1 (left) and in the adjacent quiet Sun region QS1 (middle).

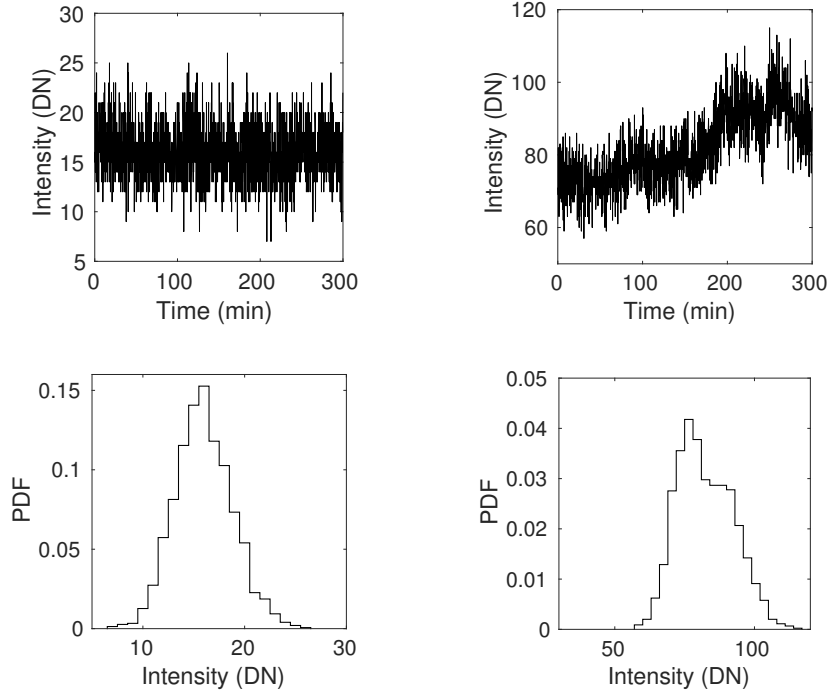

**Figure 2.** Examples of time series of 193 Å intensities (top) and corresponding probability density functions (PDFs; bottom) for single-pixel signals in the CH1 and QS1 regions. (left) A pixel in the coronal hole; (right) a pixel in the neighboring quiet-Sun region. The mean, kurtosis and skewness of the PDFs are: 16.0, 3.12, 0.18 (CH); 81.9, 2.55, 0.32 (QS).

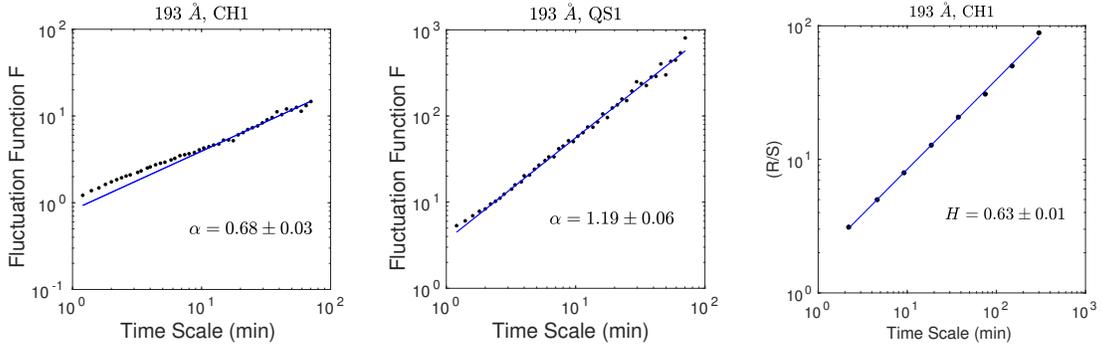

**Figure 3.** Fluctuation function $F$ as a function of temporal scale for a typical pixel in CH1 (left) and QS1 (middle). The generalized Hurst exponent $\alpha$ is obtained from the linear fits in the range $\sim 15 - 60$ min. $R/S$ as a function of temporal lag for the same pixel in CH1 (right). The Hurst exponent $H$ is calculated from the linear fit in the range $\sim 19 - 75$ min to overlap with the range for $\alpha$. The scaling symmetry actually extends from $\sim 2$ to $\sim 300$ min as shown in the figure .

The slopes of linear fits in the range $\sim 15 - 60$ min correspond to generalized Hurst exponents $\alpha = H = 0.68 \pm 0.03$ (CH1 pixel) and $\alpha = 1.19 \pm 0.06$ (QS1 pixel). The accuracy of the linear fit varies from pixel to pixel and according to the fitting range selected. We found that the long temporal range $\sim 20 - 50$ min gave a reasonable general fit for the collection of pixels. We applied the R/S analysis to the same pixel in CH1 and generated a log-log plot of $(R/S)$ vs. temporal scale (Figure 3, right). The slope led to an estimate of a Hurst coefficient $H = 0.63 \pm 0.01$ in the scaling range $\sim 19 - 75$ min (i.e., 300 min times $1/2^4, 1/2^3, 1/2^2$), which overlaps with the range used in the fit to the fluctuation function.

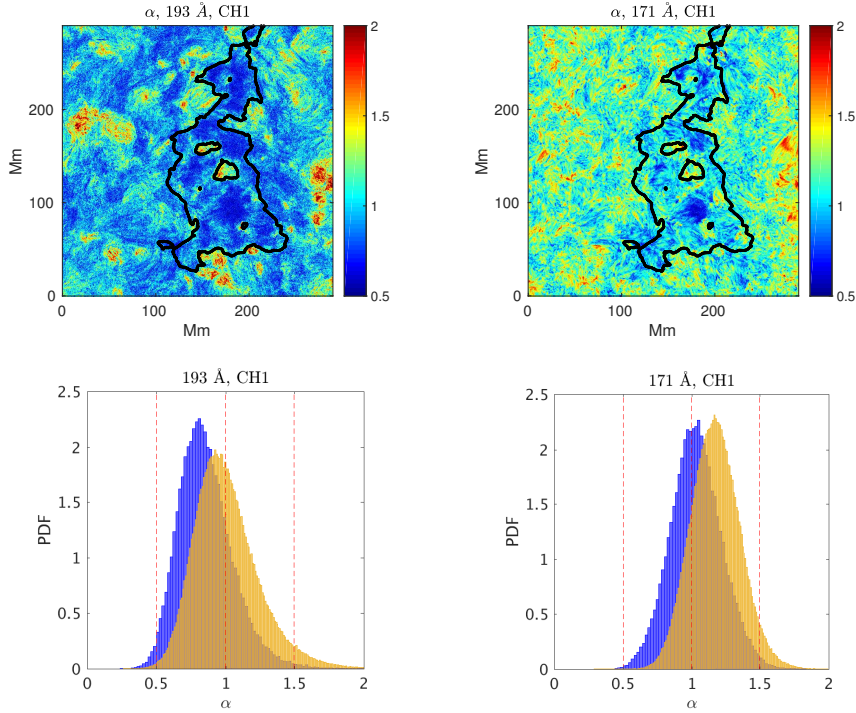

**Figure 4.** (Top row) $\alpha$ maps for the regions containing CH1 (black contour) obtained from the 193 Å (left) and 171 Å (right) time series. (Bottom row) PDFs for the $\alpha$ values corresponding to the maps in the top row (CH1 in blue, QS1 in orange).

### 4.2. *Maps and Histograms for the Generalized Hurst Exponent $\alpha$*

Applying DFA to the three regions including the CHs and QS, we found that the collection of $\alpha$ values obtained via linear fits had standard deviations with the following general properties: mean $\sim 0.05$, min $\sim 0.01$, max $\sim 0.20$. Less than 1% of the standard deviations had values $> 0.10$. The top rows of Figures 4, 5 and 6 display $\alpha$ maps for the 193 and 171 Å time series for the three regions, with the CH boundaries determined from the intensity thresholds superimposed. The bottom rows of the three figures present the PDFs of $\alpha$ values corresponding to the top rows. A summary of percentages and average values are given in Tables 1 and 2. There is large consistency in the average $\alpha$ values between the regions for the four different "classes" of pixels: persistent ($0.5 \leq \alpha \leq 1$) CH, anti-persistent ($1 < \alpha \leq 1.5$) CH, persistent QS, and anti-persistent QS. For the 193 Å data in Table 1 the $\alpha$ averages over the results for the three regions within a class are: persistent CH (0.81), anti-persistent CH (1.15), persistent QS (0.86), anti-persistent QS (1.18). In the case of the 171 Å (Table 2) signals we get: persistent CH (0.86), anti-persistent CH (1.16), persistent QS (0.91), anti-persistent QS (1.21). The standard deviations for these averages are in the order of 0.1. To fully characterize the persistent signals we have calculated the values of intensity increments and find that they satisfy an approximately Gaussian distribution (not shown here). While the average generalized Hurst exponents for the anti-persistent signals are in the range $\sim 1.15 - 1.21$, the individual values span the whole range between 1 and 1.5, as shown in the PDFs in Figures 4, 5 and 6. It is possible to identify characteristic Hurst exponents for some turbulence models via a relation with the second-order structure function $H = \xi(2)/2$ (e.g., Davies et al. 1994). Some examples are: 0.33 (Kolmogorov), 0.35 (She-Leveque), 0.37 (Müller-Biskamp). Cadavid et al. (2016) found that 171 Å emission in two hot coronal loops in an active region core had generalized Hurst exponents of 1.34 and 1.41 ($H = 0.34, 0.41$), which together with other diagnoses suggested a turbulent process. In the present case the wide range of values in the generalized Hurst exponent point toward a reconnection process. We will elaborate on this point in the discussion section. About 1 - 3 % of the pixels have $\alpha > 1.5$, which indicates a persistent, positively-correlated process also. However, in contrast to the persistent signals previously encountered, in this case the values of the intensity increments satisfy a non-Gaussian distribution. These will not be further discussed in the present study.

| Region | Percentage $0.5 \leq \alpha \leq 1$ 193 Å (%) | Mean $\alpha$ $0.5 \leq \alpha \leq 1$ 193 Å | Percentage $1 < \alpha \leq 1.5$ 193 Å (%) | Mean $\alpha$ $1 < \alpha \leq 1.5$ 193 Å | Mean $\alpha$ All Region 193 Å | Flux Imbalance |
|---|---|---|---|---|---|---|
| CH1 | 81 | 0.79 | 18 | 1.12 | 0.84 | 0.29 |
| CH2 | 67 | 0.80 | 31 | 1.16 | 0.92 | -0.29 |
| CH3 | 61 | 0.83 | 38 | 1.16 | 0.96 | -0.21 |
| QS1 | 54 | 0.84 | 43 | 1.17 | 1.01 | 0.13 |
| QS2 | 39 | 0.87 | 54 | 1.20 | 1.10 | -0.18 |
| QS3 | 41 | 0.88 | 57 | 1.17 | 1.06 | -0.06 |

**Table 1.** Statistics of the values of the generalized Hurst exponent in the 193 Å maps and the flux imbalance in the coronal holes and neighboring quiet-Sun regions.

As seen in the $\alpha$ map strengths and the PDF ranges, the difference between the CH and QS regions is based on the relative number of pixels of the two type of signals as summarized in Tables 1 and 2. The majority ($\sim 70\%$) of the 193 Å signals are persistent in the CHs. In contrast, the majority ($\sim 80\%$) of the 171 Å signals are anti-persistent in the QS regions. For the other two combinations of emission line and region type, the results are slightly more nuanced. The 193 Å emission in the QS1 region has a slight majority of persistent signals, while QS2 and QS3 have a slight majority of anti-persistent signals. In the case of the 171 Å emission, CH1 and CH3 have a slight majority of anti-persistent pixels, while CH2 has comparable amounts of the two types of signals. We note that while the intensity images for CH1 and CH3 did not present clear boundaries, the $\alpha$ maps show a better distinction between the coronal holes and the neighboring regions. For each of the three cases studied, the sub-regions inside and outside the coronal holes with magnetic bipoles have signals with $1 < \alpha \leq 1.5$.

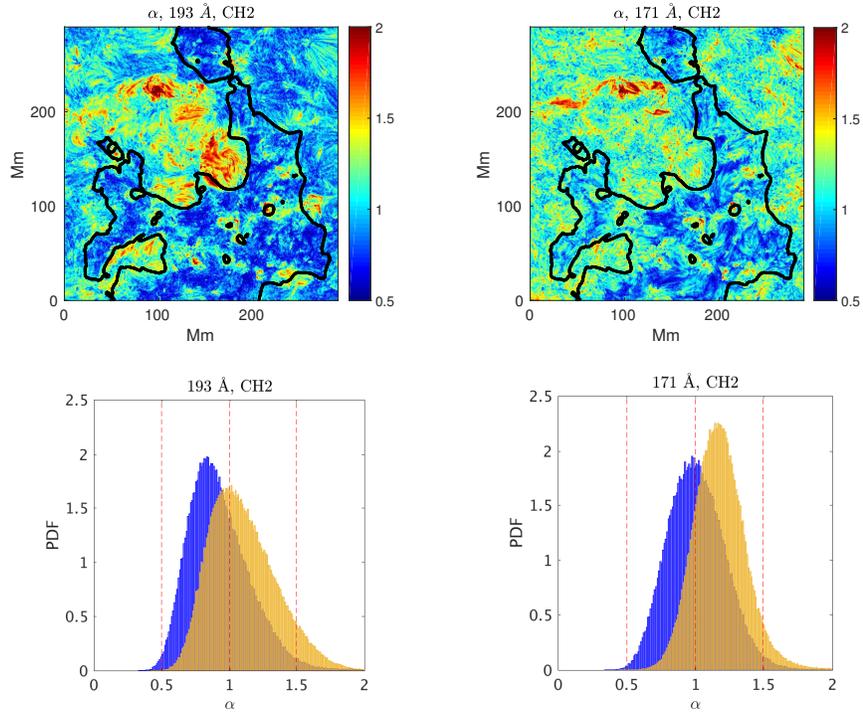

**Figure 5.** (Top row) $\alpha$ maps for the regions containing CH2 (black contour) obtained from the 193 Å (left) and 171 Å (right) time series. (Bottom row) PDFs for the $\alpha$ values corresponding to the maps in the top row (CH2 in blue, QS2 in orange).

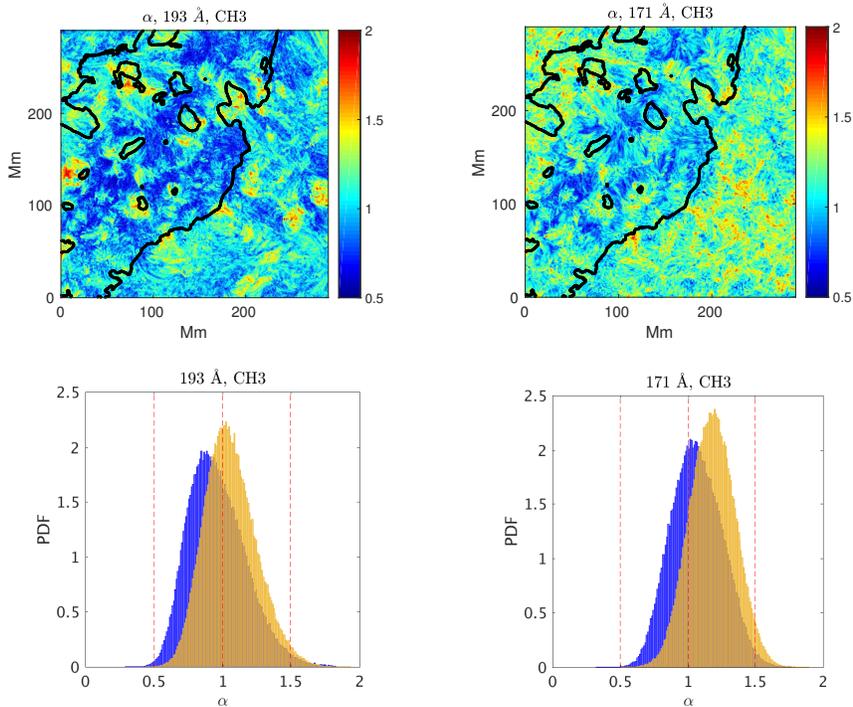

**Figure 6.** (Top row) $\alpha$ maps for the regions containing CH3 (black contour) obtained from the 193 Å (left) and 171 Å (right) time series. (Bottom row) PDFs for the $\alpha$ values corresponding to the maps in the top row (CH3 in blue, QS3 in orange).

| Region | Percentage $0.5 \leq \alpha \leq 1$ 171 Å (%) | Mean $\alpha$ $0.5 \leq \alpha \leq 1$ 171 Å | Percentage $1 < \alpha \leq 1.5$ 171 Å (%) | Mean $\alpha$ $1 < \alpha \leq 1.5$ 171 Å | Mean $\alpha$ All Region 171 Å | Flux Imbalance |
|---|---|---|---|---|---|---|
| CH1 | 45 | 0.87 | 53 | 1.15 | 1.03 | 0.29 |
| CH2 | 50 | 0.85 | 48 | 1.16 | 1.01 | -0.29 |
| CH3 | 40 | 0.87 | 58 | 1.17 | 1.07 | -0.21 |
| QS1 | 16 | 0.91 | 80 | 1.20 | 1.17 | 0.13 |
| QS2 | 16 | 0.90 | 79 | 1.21 | 1.18 | -0.18 |
| QS3 | 15 | 0.92 | 82 | 1.21 | 1.18 | -0.06 |

**Table 2.** Statistics of the values of the generalized Hurst exponent in the 171 Å maps and the flux imbalance in the coronal holes and neighboring quiet-Sun regions.

### 4.3. *Maps and Histograms Obtained with Rescaled Range Analysis*

Since the DFA of the 193 Å time series identified a majority of pixels with $\alpha = H < 1$ in the coronal holes, we proceeded to apply the R/S analysis with the expectation that it will confirm the DFA results on pixels of this type only. The top row in Figure 7 displays the $H$ maps for the three data sets. The standard deviations in the linear fits have the following general properties: mean $\sim 0.06$, min $0.00$, max $\sim 0.30$. About 15% of the pixels have a standard deviation $> 0.10$. The linear fit is performed over a range with three temporal scales only. Even with this limitation, the maps show the difference between the coronal holes and the quiet Sun regions where the method is well-defined. The percentage of pixels with mean Hurst exponents in the range $0.5 \leq H \leq 1$ are: 75%, mean 0.83 (CH1); 72%, mean 0.83 (CH2); 66%, mean 0.85 (CH3). For comparison, in the quiet Sun regions the respective values are: 41%, mean 0.89 (QS1); 33%, mean 0.90 (QS2); 43%, mean 0.90 (QS3). These results are compatible with those obtained using DFA (Table 1, columns 1 and 2). The consistency of the $H$ values between the three coronal holes is also apparent in the PDFs for all CH pixel values (Figure 7, middle row). The PDFs obtained with DFA and displayed in the bottom rows of Figures 4, 5 and 6 also show a close resemblance among the three coronal holes for $0.5 \leq \alpha \leq 1$, but they differ moderately for $\alpha > 1$.

Since R/S analysis is well-defined only for $0 \leq H \leq 1$, we have used this constraint to select pixels in the coronal holes and their boundaries with values of the Hurst exponent in the range $0.995 \leq H \leq 1.000$. The bottom row of Figure 7 shows the locations of these points superimposed on the average 193 Å intensity images for the three coronal holes. In all cases there is a tendency for these $H \sim 1$ pixels to be located near the external and internal boundaries of the coronal holes and some network lanes. The power scaling exponent at these sites is then $\beta = 2H - 1 \sim 1$, which corresponds to a $1/f$ type process.

### 4.4. *The Underlying Magnetic Field*

To quantify the relation between the values of the scaling exponents and the characteristics of the underlying magnetic field, we have calculated the flux imbalance $\langle B_z \rangle / \langle |B_z| \rangle$ in the photosphere, which gives a measure of the relative contribution of the open field (Tables 1 and 2). The CH1 region is dominated by positive polarity, while CH2 and CH3

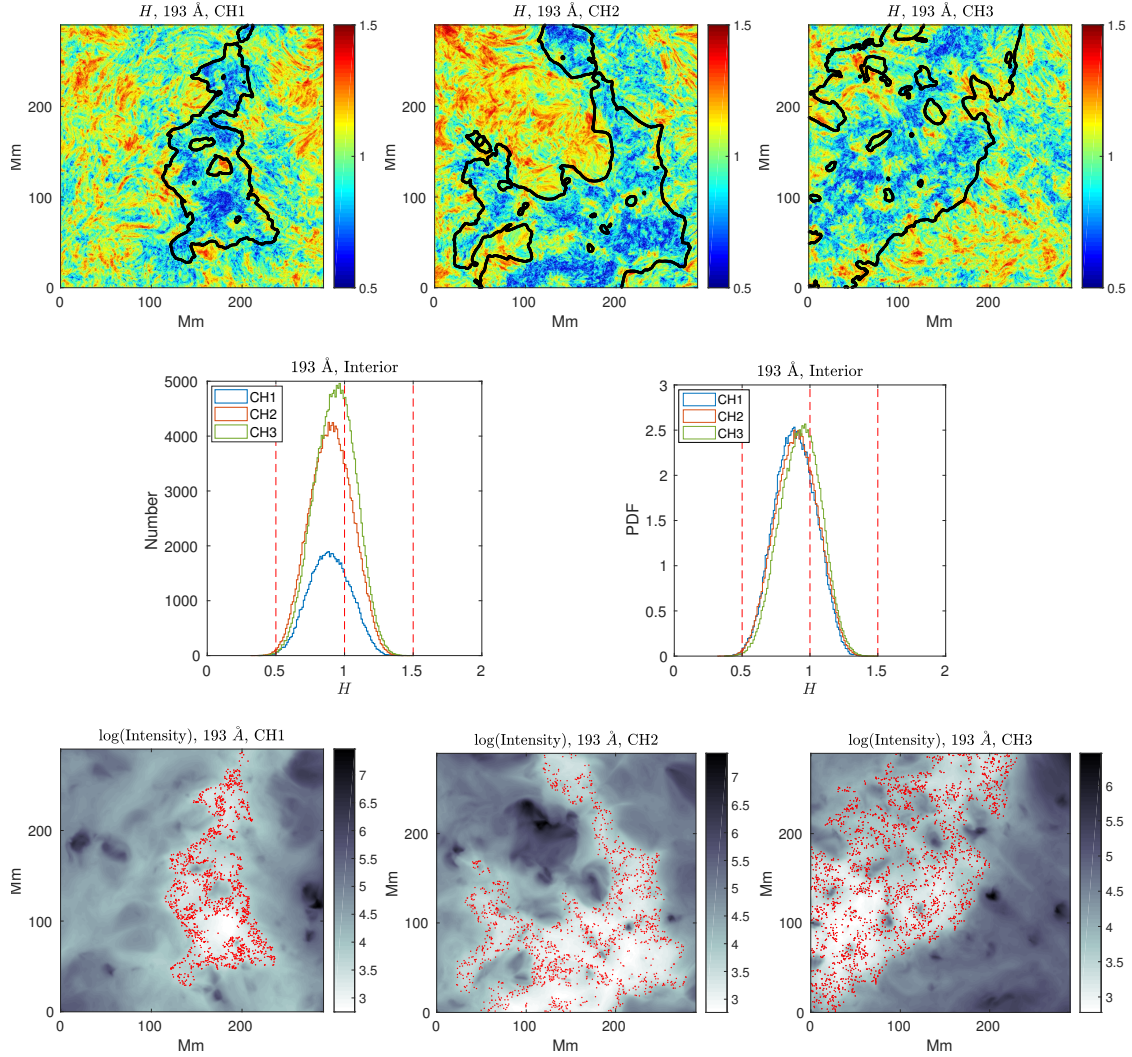

**Figure 7.** (Top row) Hurst exponent maps for the 193 Å signals in each of the three coronal hole regions (black contour) obtained using R/S analysis. This method is only valid for cases with $H \leq 1$; at other sites $H$ takes values greater than 1, which are not well-defined. (Middle row) Histogram (left) and PDF (right) of $H$ values for the interior pixels in the three coronal holes. (Bottom row) Average intensity images with $H \sim 1$ pixels in red. The grayscale has been inverted to better visualize these pixels.

have a dominant negative polarity. As expected, the flux imbalance is larger in the coronal holes, with an average absolute value of $0.26 \pm 0.05$ compared to a $0.12 \pm 0.06$ average in the neighboring quiet Sun. We have further investigated these relations in particular sub-regions. Figure 8 (top row) displays the photospheric line-of-sight magnetic field. In each case we identify three types of characteristic sub-regions which provide useful information on the interplay between flux imbalance, average scaling exponents, and average intensity in the 193 Å and 171 Å channels. The relevant quantities are summarized in Table 3. Sub-regions of type A, found within the coronal holes, are characterized by a larger flux imbalance, with an average of 0.39 for the three coronal holes. The average intensities within these sub-regions are 24 DN and 73 DN for the 193 Å and 171 Å signals, respectively; the

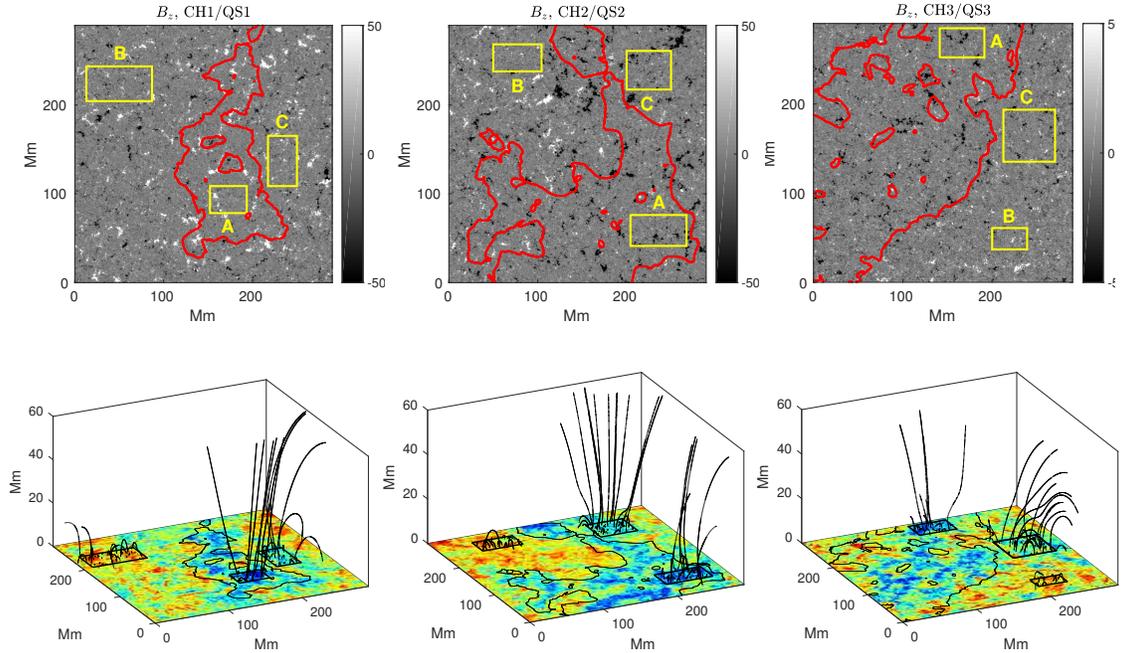

**Figure 8.** (Top row) Line-of-sight photospheric magnetic field for the three coronal hole regions: CH1 (left), CH2 (middle), CH3 (right). The field values are saturated at ±50 Gauss to accentuate the contrast. Red lines denote the coronal hole boundary and yellow boxes denote the three types of regions used to calculate the statistics presented in Table 3. (Bottom row) $H$ maps and potential magnetic field extrapolations of the photospheric fields in the top row. Regions A in the CHs are characterized mostly by open fields; regions B in the QS are dominated by closed fields; regions C in the QS present a mixture of closed loops (short and long, which appear open locally) and open-field lines.

Hurst exponent calculated using both DFA and R/S analysis is in the range $0.5 < \alpha < 1.0$. In contrast, sub-regions of type B in the neighboring quiet Sun are characterized by an average flux imbalance of 0.05 (indicating the presence of closed loop structures), higher average intensities (128 DN in the 193 Å channel and 131 DN in the 171 Å channel), and average generalized Hurst exponent $\alpha$ slightly greater than 1 (for 193 Å) and $1 < \alpha < 1.5$ (for 171 Å). Sub-regions of type C, also found in the quiet Sun, are interesting because while the flux imbalance is still large (average 0.21), indicating a larger contribution from the open field, the average intensities are higher (122 and 177 DN in the 193 Å and 171 Å channels, respectively), which should correspond to emission from closed loop structures. The average generalized Hurst exponent is 0.9 for the 193 Å signals, which we have associated with the open-field regions, and 1.10 for the 171 Å signals, which is compatible with the signals in closed-field regions.

To investigate the three-dimensional structure we have calculated potential field extrapolations of the photospheric line-of-sight (LOS) magnetic field using the Fast Fourier method of Alissandrakis (1981) and implemented with the routine `lff_extrap.pro` by M.K. Georgoulis. This choice of a current free potential field extrapolation is justified on the basis that the effects of currents are expected to have significant effects in long ($\sim 100$ Mm) coronal loops in active regions but not in the open-field coronal holes and the short loops encountered in this study (Wielgelmann & Solanki 2004).

| | Flux Imbalance | Mean Intensity (DN), 193 Å | Mean $\alpha$, 193 Å | Mean $H$, 193 Å | Mean Intensity (DN), 171 Å | Mean $\alpha$, 171 Å |
|---|---|---|---|---|---|---|
| **Sub-Regions in Coronal Holes (Type A)** | | | | | | |
| CH1 | 0.33 | 20 | 0.75 | 0.76 | 58 | 0.89 |
| CH2 | -0.36 | 22 | 0.79 | 0.77 | 53 | 0.88 |
| CH3 | -0.48 | 31 | 0.80 | 0.80 | 108 | 0.93 |
| **Sub-Regions in Quiet Sun (Type B)** | | | | | | |
| QS1 | 0.06 | 107 | 0.92 | N/A | 105 | 1.21 |
| QS2 | 0.06 | 154 | 0.98 | N/A | 177 | 1.16 |
| QS3 | 0.00 | 121 | 1.12 | N/A | 112 | 1.22 |
| **Sub-Regions in Quiet Sun (Type C)** | | | | | | |
| QS1 | 0.14 | 97 | 0.88 | 0.91 | 216 | 1.10 |
| QS2 | -0.27 | 143 | 0.88 | 0.90 | 145 | 1.12 |
| QS3 | -0.18 | 80 | 1.00 | 0.95 | 143 | 1.18 |

**Table 3.** Flux imbalance and average values for the intensities and scaling exponents in the regions defined in Figure 8.

Figure 8 (bottom row) shows the results of the extrapolation for the LOS fields in the three types of sub-regions. The quiet-Sun sub-regions of type B, with anti-persistent average $1 < \alpha \leq 1.5$ and high intensity in both the 193 Å and 171 Å channels, are primarily populated by short closed loops. For the sub-regions of types A and C (in the coronal holes and the quiet Sun, respectively), we find a mixture of short closed loops and open fields, along with partial views of large closed loops, given the curvature of the lines. For the 193 Å emission both types A and C have a persistent signature (average $\alpha$ between 0.5 and 1), but the intensity is weaker in the former. There is a predominance of open fields in sub-region type A, while type C has a larger contribution from long closed loops that appear open locally. In the case of the 171 Å signals, sub-regions of type A have a persistent signature and medium intensity, while type C have the persistent but higher-intensity signals that accompany closed short loops. Regarding the more complex results we add a note of caution that the emission from coronal holes can be contaminated by hotter plasma in the foreground, especially when large-scale close-field lines over-arch coronal holes. We will elaborate on these results in the next section.

## 5. SUMMARY AND DISCUSSION OF RESULTS

The application of DFA and R/S analysis has allowed us to identify scaling symmetries of EUV emission time series in an approximate scaling range of $\sim 20 - 50$ min. In turn, the scaling exponents provide information on the dynamics of individual pixels in three low-latitude coronal holes and their neighboring quiet-Sun regions. Although only three cases were considered, it is possible to identify characteristic patterns distinguishing the two types of regions and the boundaries.

We have found that there are three types of impulsive (bursty) signals throughout the regions considered. One is stationary and persistent with Hurst exponent $0.5 \leq H \leq 1$ (average in the range $\sim 0.8 - 0.9$). In this case the time signals, as well as the time series of their increments, tend to be Gaussian. These properties taken together suggest a fGn process. The second type of signals is nonstationary and anti-persistent with generalized Hurst exponent $1 < \alpha \leq 1.5$ and an average in the range $\sim 1.15 - 1.20$ (corresponding to an average Hurst exponent of $\sim 0.15 - 0.20$). These time series are non-Gaussian, and the time series of their increments have both Gaussian and non-Gaussian properties. For this case, the results are compatible with the properties of $1/f$ noise models. Finally, about $\sim 3\%$ of the pixels have $1.5 < \alpha \leq 2$, which corresponds to $0.5 < H \leq 1$ and indicates positive correlation. In this case neither the time signals nor the series of increments have Gaussian PDFs, distinguishing them from the other persistent time series. We do not consider this latter case in the general statistical analysis because of the low number of pixels found.

The differences between CH and QS regions in the $\alpha$ and $H$ maps comes about because of the relative number of persistent vs. anti-persistent pixels. The coronal hole regions with a higher magnetic flux imbalance have a majority of pixels ($\sim 70 - 80\%$) with low-intensity persistent 193 Å emission and medium-intensity 171 Å emission with a comparable number of pixels displaying persistent and anti-persistent properties. For the 193 Å emission the PDFs of $H$ values in the three coronal holes are remarkably similar, which if not for the non-trivial correlations would suggest an instrumental artifact. In the quiet-Sun regions with low magnetic flux imbalance, the 193 Å intensity is higher than in the CHs, and there are a comparable number of persistent and anti-persistent time series. Here the strong 171 Å emission has a majority ($\sim 80\%$) of pixels with anti-persistent signals. For the 193 Å emission in the QS, we have identified sub-regions with a persistent average Hurst exponent and a magnetic flux imbalance similar to that in the CH. However, the intensity in these regions is higher than in the CHs, leading to their classification as part of the QS using the intensity threshold criteria. We have also encountered a tendency for $H \sim 1$ near network lanes and the CH boundaries; this is compatible with $1/f$ red noise signals.

These results taken together converge to form a picture with a background weak-intensity persistent signal that can be observed in the 171 Å and 193 Å channels in both the low-density CHs and the higher-density QS regions. As previously described by various authors (e.g. Wiegelmann & Solanki 2004; McIntosh et al. 2007; He et al. 2010) and also identified in the present potential field extrapolations, short closed loops are prevalent in the QS regions but also appear in the CHs. Magnetic reconnection or oscillatory MHD fluctuations in the loops (Marsch et al. 2006a; Cranmer 2018) can cause impulsive heating events, which in turn lead to a cooling process with strong 171 Å emission anti-persistent signals and a weaker 193 Å emission at coronal temperatures. Since the number of loops is higher in the QS, the anti-persistent 171 Å emission dominates in these regions, while in the CHs there is a comparable number of anti-persistent and background persistent signals. For the 193 Å emission, except in the sub-regions with magnetic dipoles, the CH anti-persistent signals are weak and low in number and the $\alpha$ maps are dominated by the persistent background signal at the sites of weak intensity and open fields. In contrast, for the the 193 Å emission in the QS there are comparable numbers of persistent and anti-persistent signals. Generally these results confirm those of Marsch et al. (2006b), who found polar coronal hole regions with strong emission associated with small closed loops at low heights, and weak emission associated with locally open fields and originated at higher levels. The Hurst exponent allows us to probe deeper since we find sub-regions with a majority of locally open fields characterized by the persistent signal, but with intensity stronger than in the CHs as defined by the threshold technique. This suggests a direction for future research to investigate alternative approaches to defining the CH boundaries that also involve the magnetic field properties. Finally, many authors have identified interchange reconnection between open fields and closed loops in the network lanes and coronal hole boundaries (e.g. McIntosh et al. 2007; Aiouaz 2008; He et al. 2010; Subramanian et al. 2010). Given the tendency for $H \sim 1$ in these locations, we suggest that a 1/f noise signal could be associated with these processes.

The picture we are proposing here has been derived from the power laws obeyed by the EUV emission. A true test of how the underlying physics relates to the scaling parameters could only be obtained by comparing to outputs of more realistic simulations. While we do not have access to such numerical models here we first discuss how a simple phenomenological model for impulsive heating can in principle be related to emission processes with the observed scaling exponents. We follow with a description of previous work which relates observed scaling laws to models for MHD turbulence. An example in which the principles of a $1/f$ or shot noise process are applied to solar data was originally introduced by Bak et al. (1988) and elaborated upon by Aschwanden (2011) to calculate the power spectra of avalanches. Ireland et al. (2015) have further suggested its possible relevance to impulsive heating in a nanoflare model (Parker 1988; Klimchuk 2015, and references therein). The basic components of the model as are follows: An impulsive heating event with energy $E$ and time scale $T$ is given by $h(t) = (E/T) \exp(-t/T)$. It is further assumed that events satisfy an energy distribution $N \propto E^{-\alpha_E}$ with the energy and time scales related by $E \propto T^{-(1+\gamma)}$, where $\alpha_E$ and $\gamma$ are to be determined from observations or simulations. Then the total power of all events along the line of sight is given by $P(f) \propto f^{-\beta}$ with $\beta = 2\alpha - 1 = (2 - \alpha_E)(1 + \gamma)$, where $\alpha$ is the generalized exponent. An important component of the model is the relation between the energy of the impulses and the time scales. This introduces correlations in the signals, which is at the center of what our analysis has uncovered.

A more complex model based on impulsive heating by Pauluhn & Solanki (2007) was used to interpret the multiscaling properties of intensity fluctuations of the AIA 171 Å signals in an active region core and coronal loops (Cadavid et al. 2016). In this case the intensity time series result from a sequence of "nanoflares" (exponential impulses as described above), which occur at each observational time step with a certain probability $P_f$, a time scale $T$, and an amplitude $E$ sampled from a power law distribution with scaling exponent $\alpha_E > 2$. The values of the parameters where chosen to obtain probability distributions of intensities and scaling laws, comparable to those of the observations. The optimum model was characterized by a generalized exponent $\alpha = 1.20 \pm 0.02$, which covers the range of average values encountered here for the quiet Sun regions (Table 2). In this work it was also found that the ohmic dissipation rate time series resulting from a three-dimensional reduced magnetohydrodynamic (RMHD) simulation of coronal loop dynamics (Rappazzo et al. 2008), had $\alpha = 1.15 \pm 0.04$ again consistent with the values encountered here. The importance of this latter result is that there is no prescribed impulsive process but the scaling properties in the time signals result from the development of the turbulent nonlinear dynamics.

A different approach to understand the connection between scaling laws and impulsive events has been pursued by Uritsky et al. (2013). These authors considered one month of quiet-Sun magnetograms obtained with the Michelson Doppler Imager (MDI) onboard the Solar and Heliospheric Observatory (SOHO), together with simultaneous images from the Extreme Ultraviolet Images (EUVI) on the Solar TErrestrial RElations Observatory (STEREO). By setting thresholds on the values of the LOS magnetic field and the EUV emission, they identified intermittent events in the contemporaneous images of the photosphere and corona. While there is no obvious correlation on a single event basis, the analysis of the ensemble of events led to compatible scaling laws of the probability density distributions which suggested a "stochastic coupling" between the photosphere and corona. In this picture, the energy stored in the magnetic-field structures, resulting from the effects of photospheric flows, are released through a chain of "multiple spatially localized instabilities, as in the nanoflare heating model with the exception that in Uritsky et al. (2013) the thresholds were set at higher levels than the nanoflare energy scales. They found that the probability density distribution of event intensities had a scaling law compatible with that for the energy dissipation in externally driven MHD turbulence models for coronal loops (Dmitruk & Gomez 1997; Buchlin & Velli 2007). They concluded that while they did not have a full understanding of the underlying physical mechanism for the interaction between the photosphere and the corona, they expected that based on the scaling relations it would include MHD turbulence, and possibly some aspects of self-organized criticality. This complex picture is compatible with the results uncovered by our study: scaling exponent values in quiet-Sun regions characteristic of turbulent processes, and $1/f$ type signals in coronal-hole boundaries which are characteristic of self-organized criticality and shot-noise models.

## 6. ACKNOWLEDGEMENTS

We thank the referee for the useful and constructive advice which has greatly improved the paper. We appreciate the NASA/SDO and the AIA & HMI science teams for making the data available. A.C.C. acknowledges support from the Interdisciplinary Research Institute for the Sciences (IRIS) at California State University, Northridge. M.P.M. is supported by NASA grant NNX17AI27G to the Smithsonian Astrophysical Observatory. We thank Peter

Jennings for his input on the development of an efficient DFA code. We are grateful to John K. Lawrence for his useful input in the initial stages of this project.